\RequirePackage{fix-cm}
\documentclass[smallextended]{svjour3}       % onecolumn (second format)
\smartqed  % flush right qed marks, e.g. at end of proof
\usepackage{graphicx}
\begin{document}

\title{Singularity crossing, transformation of matter properties and the  problem of 
parametrization in  field theories.
}
%\subtitle{Do you have a subtitle?\\ If so, write it here}

\titlerunning{Singularity crossing, transformation of matter properties, field parametrization}        % if too long for running head

\author{A.Yu. Kamenshchik}

%\authorrunning{Short form of author list} % if too long for running head

\institute{A.Yu. Kamenshchik \at Dipartimento di Fisica e Astronomia, Universit\`a di Bologna and INFN, via Irnerio 46, 40126 Bologna, Italy\\   
              and L.D. Landau Institute for Theoretical Physics of the Russian Academy of Sciences, Kosygin street 2, 119334 Moscow, Russia\\
                            \email{kamenshchik@bo.infn.it}           %  \\
%             \emph{Present address:} of F. Author  %  if needed
}

\date{Received: date / Accepted: date}
% The correct dates will be entered by the editor

\maketitle

\begin{abstract}
We investigate particular cosmological models, based either on tachyon fields 
or on perfect fluids, for which soft future singularities arise in a natural way. 
Our main result is the description of a smooth crossing of the soft singularity 
in models with an anti-Chaplygin gas or with a particular tachyon field in the
 presence of dust. Such a crossing is made possible by certain transformations 
of matter properties. We discuss and compare also different approaches to 
the problem of crossing of the Big Bang Ð Big Crunch singularities.
% \PACS{PACS code1 \and PACS code2 \and more}
% \subclass{MSC code1 \and MSC code2 \and more}
\end{abstract}

\section{Introduction}
\label{intro}

As is well known, General relativity connects the geometrical properties of the spacetime to its matter 
content: matter tells  spacetime how to curve itself, the spacetime geometry tells  matter how to move.
The problem of cosmological singularities has been attracting the attention of theoreticians
working in gravity and cosmology since the early 1950s \cite{Lif-Khal}. In the 1960s, general theorems
about the conditions for the appearance of singularities were proven \cite{Pen-Hawk,Pen} and the oscillatory
regime of approaching the singularity \cite{BKL}, called also ``Mixmaster universe'' \cite{Misner}, was discovered.
Basically, until the end of 1990s almost all discussions about singularities were devoted to the
Big Bang and Big Crunch singularities, which are characterized by a vanishing cosmological
radius.

However, kinematical investigations of Friedmann cosmologies have raised the possibility
of sudden future singularity occurrence \cite{sudden}, characterized by a diverging  acceleration $\ddot{a}$, whereas both the
scale factor $a$ and  $\dot{a}$  are finite. Then, the Hubble parameter $H = \frac{\dot{a}}{a}$  and the energy density $\rho$
are also finite, while the first derivative of the Hubble parameter and the pressure $p$ diverge.
Until recent years, however, the sudden future singularities attracted rather a limited interest
of researchers.

The discovery of  cosmic acceleration \cite{cosmic,cosmic1} stimulated the development of ``exotic'' cosmological models of dark energy \cite{dark};  in some of these models  soft future singularities arise quite naturally.
Thus, during the last decade a plenty of work, devoted to the study of future singularities was done (see, for example, the review \cite{my-review} and references therein). 
A distinguishing feature of these singularities is the fact that rather often they can be crossed \cite{Fern,Fern1}. Indeed, due to finiteness of the Christoffel symbols, the geodesic equations are not singular and, hence, can be continued through these singularities without difficulties.   
Here a new and interesting phenomenon arises:
in some models interplay between the geometry and  matter forces  matter to change some of its basic properties, such as the equation of state for fluids and even the form of  the Lagrangian \cite{paradox1,my-review}.
Here, we consider some particular toy models, revealing this enigmatic effect.

One of the natural candidates to play the role of dark energy are the so called tachyon models, inspired by some implications of the string models 
\cite{Sen,Padman,Feinstein}. It would be more correct to call them ``Born-Infeld type'' models \cite{Born-Infeld}. A distinguishing feature of these models is a non-polynomial structure of the kinetic term in the Lagrangian. Indeed, it contains a square root. When these models are put into a cosmological context, the pressure component of the energy-momentum tensor is negative, as it should be for dark energy.  We shall use the toy tachyon model, proposed in 2004 \cite{we-tach}. It  has two particular features:
the tachyon field transforms itself into a pseudo-tachyon field;
The evolution of the universe can encounter a new type of singularity - the Big Brake singularity. The Big Brake singularity is a particular type of the soft cosmological singularities - 
the radius of the universe (i.e. scale factor) is finite, the velocity of expansion is equal to zero, the deceleration is infinite. It is the study of the models with the Big Brake singularity and 
their generalizations with the inclusion of dust-like matter, which leads us to some paradoxes and to searches of  way outs.  

Generally, the crossing of the Big Bang -- Big Crunch singularity is more complicated and looks more counter-intuitive with respect to the crossing of the soft future singularities. However, during last decade some approaches to this problem were elaborated. Behind of these approaches there are basically two general ideas. Firstly, to cross the singularity means to give a prescription matching non-singular, finite quantities before and after such a crossing. Secondly, such a description can be achieved by using a convenient choice of the field parametrization. 
Here, we present some examples of such crossing and compare briefly our approach with those recently developed \cite{Bars,Bars1,Wetterich,Wetterich1,Prester,Prester1}.   
The structure of the paper is the following: in the second section  we present  a toy tachyon model, where arises the Big Brake singularity and describe the process of its crossing; the third section is devoted to the crossing of the soft cosmological singularity in the model with the anti-Chaplygin gas and dust; in the fourth section  we consider
the crossing of the Big Brake singularity and the future of the universe in the tachyon model in the presence  of dust; fifth section is devoted to the description of the Big Bang -Big Crunch singularity crossing in the Friedmann-Lema\^{i}tre universe, filled with the scalar field; in the sixth section we consider the same problem in an anisotropic Bianchi-I universe; the last section contains some concluding remarks. 

\section{The tachyon model and the Big Brake singularity}
\label{sec:1}
We shall consider 
the flat Friedmann-Lema\^{i}tre universe
\begin{eqnarray*}
&&ds^2 = dt^2 - a^2(t)dl^2,
\end{eqnarray*}
filled with a spatially homogeneous tachyon field. 
The tachyon Lagrange density is \cite{Sen}
\begin{eqnarray*}
L = - V(T)\sqrt{1-\dot{T}^2},
\end{eqnarray*}
the energy density of the tachyon field is
\begin{eqnarray*}
\rho =  \frac{V(T)}{\sqrt{1-\dot{T}^2}},
\end{eqnarray*}
while the pressure is 
\begin{eqnarray*}
p = - V(T)\sqrt{1-\dot{T}^2}.
\end{eqnarray*}
The Friedmann equation is
\begin{eqnarray*}
H^2 \equiv \frac{\dot{a}^2}{a^2} = \rho
\end{eqnarray*}
and the equation of motion for the tachyon field is
\begin{eqnarray*}
\frac{\ddot{T}}{1-\dot{T}^{2}}+3H\dot{T}+\frac{V_{,T}}{V}=0.
\end{eqnarray*}
In our model \cite{we-tach}
\begin{eqnarray*}
&&V(T)=\frac{\Lambda }{\sin ^{2}\left[ \frac{3}{2}{\sqrt{\Lambda \,(1+k)}\ T}%
\right] }\\
&&\times\sqrt{1-(1+k)\cos ^{2}\left[ \frac{3}{2}{\sqrt{\Lambda \,(1+k)}\,T}%
\right] }\ ,  \label{VTfixed}
\end{eqnarray*}
where $k$ and $\Lambda > 0$ are the parameters of the model.
The case $k > 0$ is more interesting.
Indeed, some cosmological evolutions in the framework of this model finish in an infinite de Sitter expansion, while 
in other evolutions the tachyon field transforms into a pseudo-tachyon field
with the Lagrange density, energy density and positive pressure:
\begin{eqnarray*}
&&L = W(T)\sqrt{\dot{T}^2-1},\\
&&\rho = \frac{W(T)}{\sqrt{\dot{T}^2-1}},\\ 
&&p = W(T)\sqrt{\dot{T}^2-1},\\
&&W(T) = \frac{\Lambda }{\sin ^{2}\left[ \frac{3}{2}{\sqrt{\Lambda \,(1+k)}\ T}%
\right] }\\
&&\times\sqrt{(1+k)\cos ^{2}\left[ \frac{3}{2}{\sqrt{\Lambda \,(1+k)}\,T}-1
\right] }.
\end{eqnarray*}
Such a transformation is caused by the fact that at some point of the phase space the kinetic term and the potential term become simultaneoulsly  equal to zero and should change their signs, which makes them imaginary. At the same time the equation of motion are regular. Thus, it is necessary to redefine both the kinetic and potential terms as it was done in the formulae above \cite{we-tach}.  

What happens to the Universe after the transformation of the tachyon into the pseudo-tachyon 
? It encounters the Big Brake 
cosmological singularity \cite{we-tach}.
\begin{eqnarray*}
t \rightarrow t_{BB} < \infty,
\end{eqnarray*}
\begin{eqnarray*}
a(t \rightarrow t_{BB}) \rightarrow a_{BB} < \infty, 
\end{eqnarray*}
\begin{eqnarray*}
\dot{a}(t \rightarrow t_{BB}) \rightarrow 0,
\end{eqnarray*}
\begin{eqnarray*}
\ddot{a}(t \rightarrow t_{BB}) \rightarrow -\infty,
\end{eqnarray*}
\begin{eqnarray*}
R(t \rightarrow t_{BB}) \rightarrow +\infty,
\end{eqnarray*}
%\begin{equation*}
%T(t \rightarrow t_{BB}) \rightarrow T_{BB},\ |T_{BB}| < \infty
%\end{equation*}
%\begin{equation*}
%|\dot{T}(t \rightarrow t_{BB})| \rightarrow \infty
%\end{equation*}
\begin{eqnarray*}
\rho(t \rightarrow t_{BB}) \rightarrow 0,
\end{eqnarray*}
\begin{eqnarray*}
p(t \rightarrow t_{BB}) \rightarrow +\infty.
\end{eqnarray*}
If $\dot{a}(t_{BB}) \neq 0$ it is a 
more general soft singularity.
At the Big Brake singularity the equations for geodesics are regular, because 
the Christoffel symbols are regular (moreover, they are equal to zero).
 Thus, it shoud be  possible  to cross the Big Brake singularity. This process can be described in detail. 
On analyzing the equations of motion we find that on approaching the Big Brake singularity  the tachyon field behaves as \cite{we-tach1}

\begin{eqnarray*} 
T=T_{BB}+\left( \frac{4}{3W(T_{BB})}\right) ^{1/3}(t_{BB}-t)^{1/3}.
\label{tachBB}
\end{eqnarray*}
Its time derivative $s \equiv \dot{T}$
behaves as 
\begin{eqnarray*}
s=-\left( \frac{4}{81W(T_{BB})}\right) ^{1/3}(t_{BB}-t)^{-2/3},  \label{sBB}
\end{eqnarray*}
while the cosmological radius is 
\begin{eqnarray*}
a=a_{BB}-\frac{3}{4}a_{BB}\left( \frac{9W^{2}(T_{BB})}{2}\right)
^{1/3}(t_{BB}-t)^{4/3},  \label{cosmradBB}
\end{eqnarray*}
its time derivative is 
\begin{eqnarray*}
\dot{a}=a_{BB}\left( \frac{9W^{2}(T_{BB})}{2}\right) ^{1/3}(t_{BB}-t)^{1/3}
\label{cosmradderBB}
\end{eqnarray*}
and the Hubble variable is 
\begin{eqnarray*}
H=\left( \frac{9W^{2}(T_{BB})}{2}\right) ^{1/3}(t_{BB}-t)^{1/3}.
\label{HubbleBB}
\end{eqnarray*}
All these expressions can be continued into the region where $t>t_{BB}$,which
amounts to crossing the Big Brake singularity. Only the expression for $s$
 is singular at $t=t_{BB}$, but this singularity is integrable and
not dangerous.

Once reaching the Big Brake, it is impossible for
the universe to stay there because of the infinite 
deceleration, which eventually leads to a decrease
of the scale factor. This is because after the Big
Brake crossing the
time derivative of the
cosmological radius  and Hubble
variable  change
their signs. The expansion is then followed
by a contraction, culminating in the Big Crunch singularity.

\section{Crossing of the soft singularity in the model with the anti-Chaplygin gas and dust}
\label{sec:2}
One of the simplest cosmological models revealing a Big Brake singularity is the model based on the anti-Chaplygin gas with an equation of state 
\begin{eqnarray*}
p = \frac{A}{\rho},\ \ A > 0.
\end{eqnarray*}
Such an equation of state arises, for example,  in the theory of wiggly strings \cite{wiggly,wiggly1}.
The energy density now is 
\begin{eqnarray*}
\rho(a)=\sqrt{\frac{B}{a^6}-A}.
\end{eqnarray*}
At $a = a_* = \left(\frac{B}{A}\right)^{1/6}$ the universe encounters the Big Brake singularity.

In the universe filled by the  
anti-Chaplygin gas and dust
the energy density and the pressure are 
\begin{eqnarray*}
\rho(a) = \sqrt{\frac{B}{a^6}-A} + \frac{M}{a^3},\ \ p(a) =\frac{A}{\sqrt{\frac{B}{a^6}-A}}.
\end{eqnarray*}
Due to the presence of the dust component, the Hubble parameter has a
non-zero value at the encounter with the singularity, therefore the dust
implies further expansion. With continued expansion however, the energy
density and the pressure of the anti-Chaplygin gas would become ill-defined.

In principle, one can resolve this paradox, considering an abrupt transition from an expansion to a contraction at the finite (non-vanishing) expansion rate.
It could be done self-consistently on using  generalized functions \cite{paradox}.     
However, the abrupt transition from the expansion to the contraction of the universe 
does not look natural. There is an alternative  way of resolving the paradox.
One can try to change the equation of state of the anti-Chaplygin gas on passing the soft 
singularity \cite{paradox1,my-review}.
There is some analogy between the transition from an expansion to a contraction of a universe   and the perfectly elastic bounce of a ball from a wall in classical mechanics.
There is  also an abrupt change of the direction of the velocity (momentum).
However, we know that in reality the velocity is changed continuously 
due to the deformation of the ball and of the wall. Something similar occurs also in the cosmology. 
To see this let us discuss what happens when the anti-Chaplygin gas arrives closely to the singularity.

The pressure of the anti-Chaplygin gas 
\begin{eqnarray*}
p = \frac{A}{\sqrt{\frac{B}{a^6}-A}}  \label{pressure-anti}
\end{eqnarray*}
tends to $+\infty$ when the universe approaches the soft singularity.
Requiring the expansion to continue into the region $a > a_S$, while
changing minimally the equation of state, we  assume 
\begin{eqnarray*}
p = \frac{A}{\sqrt{|\frac{B}{a^6}-A|}},  \label{pressure-new}
\end{eqnarray*}
\begin{eqnarray*}
p = \frac{A}{\sqrt{A-\frac{B}{a^6}}},\ \mathrm{for}\ a > a_S.
\label{pressure-new1}
\end{eqnarray*}
Using the energy conservation equation 
\begin{eqnarray*}
\dot{\rho}+3\frac{\dot{a}}{a}(\rho+p)=0,
\end{eqnarray*}
we obtain
the energy density
\begin{eqnarray*}
\rho= -\sqrt{A-\frac{B}{a^6}}.
\end{eqnarray*}
The anti-Chaplygin gas transforms itself into the Chaplygin gas \cite{Chap}, 
i.e. the gas with the equation of state 
\begin{eqnarray*}
p =-\frac{A}{\rho},\ \ A > 0,
\end{eqnarray*}
with 
a negative energy density.
The pressure remains positive and the  expansion continues.
The spacetime geometry also remains continuous.  
The expansion stops at $a = a_0$, where 
\begin{eqnarray*}
\frac{M}{a_0^3} -\sqrt{A-\frac{B}{a_0^6}} = 0.
\end{eqnarray*}
Then the contraction of the universe begins. 
At the moment when the energy density of the 
Chaplygin gas becomes equal to zero 
(again we encounter a soft singularity), the Chaplygin gas transforms itself back into the anti-Chaplygin gas
 and the contraction continues culminating in an encounter with the Big Crunch singularity.
Thus, we have obtained a self-consistent picture of the cosmological evolution of the Friedmann-Lema\^{i}tre universe filled with 
the anti-Chaplygin gas and dust. 

\section{Crossing the Big Brake singularity and the future of the universe in the tachyon model in the presence of dust}
\label{sec:3}
Now we shall try to consider a more complicated model. 
What happens with the Born-Infeld type pseudo-tachyon field in the presence
of a dust component ? Does the universe still run into a soft singularity?
To see that the answer to this question is affirmative, it is enough to find the time dependence of the pseudo-tachyon field in the vicinity of the singularity: 
\begin{eqnarray*}
T=T_{S}\pm \sqrt{\frac{2}{3H_{S}}}\sqrt{t_{S}-t},\ H_{S}=\sqrt{\frac{\rho _{m,0}}{a_{S}^{3}}}.  \label{new-asymp}
\end{eqnarray*}
How can the universe cross this singularity?
To answer this question it is convenient to begin with the analysis of the simplest case: the pseudo-tachyon field with a constant potential.
A pseudo-tachyon field with a constant potential
is equivalent to the anti-Chaplygin gas. (The fact that the tachyon field with the constant potential is equivalent to the Chaplygin gas was noticed in \cite{tach-Chap}).
In this case to the change of the equation of state of the anti-Chaplygin gas corresponds the following transformation of the Lagrangian of the pseudo-tachyon field:
\begin{eqnarray*}
L=W_{0}\sqrt{\dot{T}^{2}+1}.
\end{eqnarray*}
Correspondingly, the pressure and the energy-density are 
\begin{eqnarray*}
p=W_{0}\sqrt{\dot{T}^{2}+1},  \label{pressure-BI}
\end{eqnarray*}
\begin{eqnarray*}
\rho =-\frac{W_{0}}{\sqrt{\dot{T}^{2}+1}}.  \label{energy-BI}
\end{eqnarray*}
It is  a new type of Born-Infeld field,
which we may call ``quasi-tachyon''. 

For an arbitrary potential the
Lagrangian reads
\begin{eqnarray*}
L=W(T)\sqrt{\dot{T}^{2}+1},~~~~~~a>a_S  \label{Lagr-new}
\end{eqnarray*} 
and 
\begin{eqnarray*}
\frac{\ddot{T}}{\dot{T}^{2}+1}+3H\dot{T}-\frac{W_{,T}}{W}=0,  \label{KG-BI}
\end{eqnarray*} 
\begin{eqnarray*}
\rho = -\frac{W(T)}{\sqrt{\dot{T}^2+1}},~  \label{energy-BI1}
\end{eqnarray*}
\begin{eqnarray*}
p = W(T)\sqrt{\dot{T}^2+1}.
\end{eqnarray*}

Why can we make this kind of the generalization for a Lagrangian of a Born-Infeld type field?
In the vicinity of the soft singularity the 
friction term $3H\dot{T}$ in the equation of motion  dominates over the potential term $W_{,T}/W$. Hence, the
dependence of $W(T)$ on its argument is not essential and a pseudo-tachyon field approaching
this singularity behaves like one with a constant potential. Thus, it is
reasonable to assume that upon crossing the soft singularity the pseudo-tachyon
transforms itself into a quasi-tachyon 
for any potential $W(T)$.

\section{Big Bang -Big Crunch crossing in the Friedmann-Lema\^{i}tre universe, filled with the scalar field}
\label{sec:4}

The idea that the Big Bang - Big Crunch singularity can be crossed appears very counterintuitive. However, during the last decade some approaches to the description of this crossing were elaborated \cite{Bars,Bars1,Wetterich,Wetterich1,Prester,Prester1}. Behind these approaches there is a general idea: on choosing some convenient  parametrization of the metric and of the matter fields one can provide a matching between the characteristics of the universe before and after the singularity crossing.
Here there is an analogy with the horizon which arises due to a certain choice of the spacetime coordinates: the singularity arises because of some choice of the field parametrization.
On choosing appropriate combinations of the field variables we can describe the passage through the Big Bang - Big Crunch singularity, but this does not mean that the presence of such a singularity is not essential. Indeed, extended
  objects cannot survive this passage. 
  In this sense one can trace an analogy to the Kruskal coordinates for the Schwarzschild geometry: the introduction of such coordinates does not eliminate the horizon and 
the effects, connected with its existence, but, instead, permits us to see what happens behind the horizon. 

In paper \cite{we-cross} we have considered a flat Friedmann-Lema\^{i}tre universe filled with a scalar field, non-minimally coupled with the scalar curvature.
The action in this model looks as 
\begin{eqnarray*}
S =\int d^4x\sqrt{-g}\left[U(\sigma)R - \frac12g^{\mu\nu}\sigma_{,\mu}\sigma_{,\nu}+V(\sigma)\right]
\label{action0}
\end{eqnarray*}
An interesting particular case is the case of the conformal coupling between the curvature and the scalar field when  
\begin{eqnarray*}
U(\sigma) = U_0-\frac{1}{12}\sigma^2.
\end{eqnarray*}
A conformal transformation of the metric
\begin{eqnarray*}
g_{\mu\nu} = \frac{U_1}{U}\tilde{g}_{\mu\nu},
\label{conf}
\end{eqnarray*}
toghether with the introduction of a new scalar field $\phi$: 
\begin{eqnarray*}
\frac{d\phi}{d\sigma} = \frac{\sqrt{U_1(U+3U'^2)}}{U}
\Rightarrow
\phi =\! \int\! \frac{\sqrt{U_1(U+3U'^2)}}{U} d\sigma,
\label{scal1}
\end{eqnarray*}
\begin{eqnarray*}
\phi = \sqrt{3U_1}\ln \left[\frac{\sqrt{12U_0}+\sigma}{\sqrt{12U_0}-\sigma}\right],
\label{connection_c0}
\end{eqnarray*}
\begin{eqnarray*}
\sigma = \sqrt{12U_0}\tanh\left[ \frac{\phi}{\sqrt{12U_1}}\right],
\label{connection_c}
\end{eqnarray*}
bring us to the action  for a minimally coupled scalar field:
\begin{eqnarray*}
S =\int d^4x\sqrt{-\tilde{g}}\left[U_1R(\tilde{g}) - \frac12\tilde{g}^{\mu\nu}\phi_{,\mu}\phi_{,\nu}+W(\phi)\right] ,
\label{action1e}
\end{eqnarray*}
with 
\begin{eqnarray*}
W(\phi) = \frac{U_1^2 V(\sigma(\phi))}{U^2(\sigma(\phi))}.
\label{poten}
\end{eqnarray*}
This  is called the transformation from the Jordan 
frame to the Einstein frame \cite{Wagoner}.

In a flat Friedmann-Lema\^{i}tre universe 
\begin{eqnarray*}
ds^2=N^2d\tau^2-a^2dl^2,
\end{eqnarray*}
\begin{eqnarray*}
d\tilde{s}^2 =\tilde{N}^2d\tau^2-\tilde{a}^2dl^2.
\end{eqnarray*}
The relations between the lapse  functions $N$ and $\tilde{N}$ and the scale factors $a$ and $\tilde{a}$  in these two frames are given by 
\begin{eqnarray*}
\tilde{N} = \sqrt{\frac{U}{U_1}}N,\ \tilde{a}=\sqrt{\frac{U}{U_1}}a.
\end{eqnarray*}
Besides,
\begin{eqnarray*}
t = \int \sqrt{\frac{U_1}{U}}d\tilde{t},
\end{eqnarray*}
where $t$ and $\tilde{t}$ are the cosmic time parameters in the Jordan and the Einstein frames. 
 In terms of the scalar field, the relation between the scale factors can be rewritten as
 \begin{eqnarray*}
a = \tilde{a}\sqrt{\frac{U_1}{U_0}}\cosh\left(\frac{\phi}{\sqrt{12U_1}}\right).
\end{eqnarray*}

The transitions between the Jordan frame and the Einstein frame can be used to find the solutions of the dynamical equations for some cosmological model in one frame 
if some such solutions are known for its counterpart in another frame. This method was used in papers \cite{frame,frame1}, where starting from some exact solutions obtained in \cite{Sagnotti} for models with a minimally coupled scalar field , were constructed their counterparts in the models with conformal coupling and in the induced gravity models \cite{induced}. 
There exists a lot of works devoted to the question of the so called equivalence of the Jordan and of the Einstein frames. Especially complicated looks this problem in the quantum context (see e.g. \cite{Steinwachs,Steinwachs1} and the references therein). 

We wish to emphasise that from our point of view the Einstein frame and the Jordan frame are mathematically equivalent and  self-consistent description of reality can be constructed in any frame. However, the physical cosmological evolutions are those seen by an observer using the cosmic (synchronous) time, which is different in different frames. Thus, evolutions in the Einstein and Jordan frames, connected by a conformal transformation and by the reparametrization of the scalar field can be qualitatively different. We have constructed an explicit example of such a difference in \cite{KPTVV2013}. More precisely, we have considered a de Sitter expansion in induced gravity with a scalar field squared self-interaction potential~\cite{we-ind-ex} and shown that its counterpart is the well-known particular power-law solution~\cite{exp-part} in a minimally coupled model with an exponential potential. We think that while the general solutions for some non-minimally coupled models can be obtained from their minimally coupled counterparts, the study of their behaviour is of interest because it can be physically different. The fact that the expansion in one frame can correspond to the contraction in another frame has also been noticed in the literature~\cite{Gasp,Pol1}.

Here we shall use that transitions between the frames to describe the singularity crossing. Indeed, at the moment of the arriving to the singularity in one frame, the universe can have quite regular evolution in another frame. It does not mean that the singularity disappears in some enigmatic way, it is simply shifted to the singular conformal transformation between the frames. However, we shall see that it is enough to describe the crossing of the singularity in a self-consistent way.  We shall use the simplest model of the massless scalar field with the vanishing potential because already in this case all the basic features of our approach can be seen.

In the vicinity of the singularity in the Einstein frame:
\begin{eqnarray*}
\tilde{a} \sim \tilde{t}^{\frac13} \rightarrow 0,\ {\rm when}\ \tilde{t} \rightarrow 0.
\end{eqnarray*}
However, in the Jordan frame:
\begin{eqnarray*}
a \sim \tilde{t}^{\frac13}\left(\tilde{t}^{\frac13}+\tilde{t}^{-\frac13}\right) \rightarrow {\rm const} \neq 0.
\end{eqnarray*}
Meanwhile, the scalar field $\sigma$ crosses the value $\pm\sqrt{12U_0}$ and 
the coupling function $U$ changes its sign. 
Thus, the evolution in the Jordan frame is regular, and we can use this fact to describe the 
crossing of the Big Bang - Big Crunch singularity in the Einstein frame.

If one considers the expansion of the universe from the Big Bang with normal gravity driven by the standard scalar field,
the continuation backward in time shows that it was preceded by the contraction 
towards a Big Crunch singularity in the antigravity regime, driven by a phantom scalar field 
with a negative kinetic term \cite{we-cross}.

The possibility of a change of sign of the effective gravitational constant in the model with a conformably coupled scalar field was analysed in 1981 by A. Starobinsky \cite{Star1981}, 
following the earlier suggestion made by A. Linde in 1980 \cite{Linde1980}. 
It was shown that in a homogeneous and isotropic universe, one can indeed cross the point where the effective gravitational constant changes sign. However,  the presence of anisotropies changes the situation: these anisotropies grow indefinitely when this constant is equal to zero.  In the next section we shall describe this effect in our language and we shall show why the procedure, described in this section does not work for a simple model of an anisotropic Bianchi-I universe. 

\section{Singularity crossing in a Bianchi - I universe}
\label{sec:5}

Let us consider the simplest anisotropic cosmological model, the model of a Bianchi-I universe. Using this example, we shall show that the method of the description of the crossing of the singularity, described in the preceding section and in paper \cite{we-cross} does not work in this case \cite{we-cross1}. 

The metric of the Bianchi-I universe in the Einstein frame is
\begin{eqnarray*}
d\tilde{s}^2 = \tilde{N}(\tau)^2d\tau^2-\tilde{a}^2(\tau)(e^{2\beta_1(\tau)}dx_1^2+e^{2\beta_2(\tau)}dx_2^2+e^{2\beta_3(\tau)}dx_3^2),
\end{eqnarray*}
while in the Jordan frame it is  
\begin{eqnarray*}
ds^2 = N(\tau)^2d\tau^2-a^2(\tau)(e^{2\beta_1(\tau)}dx_1^2+e^{2\beta_2(\tau)}dx_2^2+e^{2\beta_3(\tau)}dx_3^2),
\end{eqnarray*}
where
\begin{eqnarray*}
\beta_1+\beta_2+\beta_3=0.
\end{eqnarray*}
One can show that 
\begin{eqnarray*}
\dot{\beta}_i=\frac{\beta_{i0}}{\tilde{a}^3}, \ \theta_0=\beta_{10}^2+\beta_{20}^2+\beta_{30}^2.
\end{eqnarray*}
\begin{eqnarray*}
\dot{\phi} = \frac{\phi_0}{\tilde{a}^3},\ \phi = \frac{\phi_0}{\left(\frac{3\theta_0}{2}+\frac{3\phi_0^2}{4U_1}\right)^{\frac12}}\ln \tilde{t}.
\end{eqnarray*}
In the vicinity of the singularity in the Einstein frame one has as before
\begin{eqnarray*}
\tilde{a} \sim \tilde{t}^{\frac{1}{3}}.
\end{eqnarray*}
However, in the Jordan frame 
\begin{eqnarray*}
a \sim  \tilde{t}^{\frac{1}{3}}(\tilde{t}^{\gamma}+\tilde{t}^{-\gamma}) \rightarrow 0,
\end{eqnarray*}
because 
\begin{eqnarray*}
\gamma= \frac{\phi_0}{3\sqrt{\phi_0^2+2\theta_0U_1}} < \frac13.
\end{eqnarray*}
Thus, in the presence of a anisotropy one also encounters the Big Bang singularity  in the Jordan frame.

We shall look now for another way of reparametrizing of the fields, which could permit the description of the singularity crossing.
As in the case of the transitions between the Jordan and the Einstein frames, such a change of the reparametrizations involves some 
mixing between the geometrical and matter degrees of freedom. 

Let us for a moment come back to a Friedmann-Lema\^{i}tre flat cosmological model filled with a massless scalar field, minimally coupled to gravity.
The Lagrangian of this model can be represented as 
\begin{eqnarray*}
L = \frac12\dot{x}^2-\frac12\dot{y}^2,
\end{eqnarray*}
where
\begin{eqnarray*}
x = \frac{4\sqrt{U_1}}{\sqrt{3}}\tilde{a}^{\frac32} \cosh \frac{\sqrt{3}}{4\sqrt{U_1}}\phi,\ y= \frac{4\sqrt{U_1}}{\sqrt{3}}\tilde{a}^{\frac32} \sinh \frac{\sqrt{3}}{4\sqrt{U_1}}\phi,
\end{eqnarray*}
and the Friedmann equation is 
\begin{eqnarray*}
\dot{x}^2-\dot{y}^2=0.
\end{eqnarray*}
Inversely,
\begin{eqnarray*}
\tilde{a}^3 = \frac{3(x^2-y^2)}{16U_1},
\end{eqnarray*}
\begin{eqnarray*}
\phi=\frac{4\sqrt{U_1}}{\sqrt{3}}{\rm arctanh}\frac{y}{x}.
\end{eqnarray*}
This kind of the Lagrangians, including also some potential terms of a particular form was used for the construction of some exact cosmological solutions of the Einstein equations \cite{change,change1,change2,change3,change4,change5}. 
Initially 
\begin{eqnarray*}
x>|y|.
\end{eqnarray*}
The solution of the equations of motion is 
\begin{eqnarray*}
x=x_1\tilde{t}+x_0,\ y=y_1\tilde{t}+y_0,\ x_1^2=y_1^2.
\end{eqnarray*}
Choosing the constants as 
\begin{eqnarray*}
x_0=y_0=A>0,\ x_1=-y_1 = B>0,
\end{eqnarray*}
we have
\begin{eqnarray*}
\tilde{a}^3=\frac{3AB\tilde{t}}{4U_1}.
\end{eqnarray*}
Now, we can make a continuation in the plane $(x,y)$, to $x < |y|$ or, in other words, to  $\tilde{t} < 0$.
Such a continuation implies an antigravity regime and the transition to the phantom scalar field, just 
as in the more complicated schemes, discussed before.  Thus, using this very simple reparametrization of the gravitational and matter (scalar) field, we
have managed to describe the singularity crossing \cite{we-cross1}. 

How can we generalize these considerations to the case when the anisotropy term is present ?
Let us consider a system with the Lagrangian
\begin{eqnarray*}
L = \frac12\dot{r}^2-\frac{1}{2}r^2(\dot{\varphi}^2+\dot{\varphi}_1^2+\dot{\varphi_2}^2),
\end{eqnarray*}
where 
\begin{eqnarray*}
\varphi_1=\sqrt{\frac38}\alpha_1,\ \varphi_2\sqrt{\frac38}\alpha_2,
\end{eqnarray*}
\begin{eqnarray*}
\beta_1=\frac{1}{\sqrt{6}}\alpha_1+\frac{1}{\sqrt{2}}\alpha_2,\ \beta_2=\frac{1}{\sqrt{6}}\alpha_1-\frac{1}{\sqrt{2}}\alpha_2,\ \beta_3=-\frac{2}{\sqrt{6}}\alpha_1.
\end{eqnarray*}
We can again consider the plane $(x,y)$ as 
\begin{eqnarray*}
&&x = r\cosh \Phi,\\
&&y=r\sinh \Phi,
\end{eqnarray*}
where a new hyperbolic angle $\Phi$ is defined by
\begin{eqnarray*}
\Phi=\int d\tilde{t}\sqrt{\dot{\varphi}_1^2+\dot{\varphi}_2^2+\dot{\varphi}^2}.
\end{eqnarray*}

We have reduced a four-dimensional problem to the old two-dimensional one, on using the fact that the variables $\alpha_1,\alpha_2$ and $\phi$ enter into the equation of motion for the scale factor $\tilde{a}$ only through the squares of their time derivatives.
The behaviour of the scale factor before and after the crossing of the singularity can be matched by using the transition to the new coordinates $x$ and $y$, which mix geometrical and scalar field variables in a particular way.
To describe the behaviour of the anisotropic factors it is enough to fix the constants $\beta_{i0}$.    
 
 \section{Conclusions and discussion}
 \label{sec:6}

In the present work we have considered some attempts of the description of crossing of different kinds of the cosmological singularities. First, we have considered
the crossing of some future soft singularities. The fact that such singularities can be crossed is rather known and does not provoke great controversies. A particular 
feature of the models, considered in paper \cite{paradox1} and in the preceding papers is that the process of the crossing of thew singularity is described in detail and is at some circumstances accompanied by the geometrically induced transformations of matter properties. 

We would like to dwell on this fact. Concerning the transformation from the anti-Chaplygin gas  to the Chaplygin gas with negative energy density, we would like to emphasize that this is not an extension of the definition of the anti-
Chaplygin gas into the region, where it was not defined before, but instead that it is a transition from one perfect
fluid to another one. This transition is the result of a complicated interplay between the evolution of the spacetime
described by the Friedmann equations and the evolution of perfect fluids, described by the continuity
equations. Indeed, in the description of this transition we
use not only the equations of state of fluids, but also the
explicit dependences of their energy densities and pressure
on the cosmological radius. Thus, in describing the passage
of the universe filled with the anti-Chaplygin gas and with
dust through the soft singularity, we put forward two
requirements: first, the cosmological evolution should be
as smooth as possible; second, the change of the character
of the dependence of the energy density and of the pressure
of the fluids should be minimal. These two requirements
imply the transformation of the anti-Chaplygin gas into a Chaplygin gas with a negative energy density. 

The situation with transformations of the tachyon field is
more complicated. First of all, let us note that there are two
different kinds of transformations, the transformation from
tachyon to pseudo-tachyon and the transformation from
pseudo-tachyon to quasi-tachyon. The first kind of transformation
was introduced in the paper \cite{we-tach}.  This transformation is not connected
with the crossing of the singularity. When the pressure of
the tachyon field vanishes, the potential and kinetic terms
in the Lagrangian  become ill defined. However, the
equations of motion of this field can be continued to the
part of the phase space of the corresponding dynamical
system, where the pressure is positive. The new Lagrangian
well defined in this region, gives the equation
of motion which coincides with the old equation of
motion. 
The justification of the transition from the pseudo-tachyon
field to the quasi-tachyon field
 is more subtle. This transformation is induced by the
crossing of the soft singularity in the presence of dust and
there is no way to use the continuity of the form of the
Lagrangian or the conservation of the form of the equations
of motion. We use instead the fact that the equation of state
of the pseudo-tachyon field with constant potential coincides
exactly with that of the anti-Chaplygin gas. Thus, to
provide a passage which is as smooth as possible of the
universe filled with the pseudo-tachyon field with constant
potential through the soft singularity we should find such a
Lagrangian of a Born-Infeld-type field which is equivalent
to the Chaplygin gas with a negative energy density.
Following this path we come to the quasi-tachyon field.
 The last step consists in the generalization of this scheme 
for the case of an arbitrary
potential. Such a generalization is justified by the fact,
that in the vicinity of the soft singularity, the change of the
potential term of the pseudo-tachyon field is much slower in
comparison with that of the kinetic term.
While the transition from the pseudo-tachyon to the
quasi-tachyon is more radical and involved than the other
matter transformations considered here and in the preceding
papers, it still looks quite logical and probably the only
one which is possible.

As we have already said above the idea of the crossing of the Big Bang - Big Crunch 
singularity looks much more counterintuitive with respect to the crossing of soft singularities.
Nonetheless,  the procedure for the crossing of the Big Bang - Crunch singularity, based on the use of  Weyl symmetry, was elaborated \cite{Bars,Bars1}.  Using a Weyl - invariant theory, where two conformally coupled
with gravity scalar fields were presented, the authors  obtained the geodesic completeness of the corresponding spacetime. The consequence  of this geodesic completeness is the crossing of the Big Bang singularity and the emergence of  antigravity regions on using the Einstein frame. 
The use of Weyl symmetry to describe the passage through the Big Crunch - Big Bang singularity accompanied by a change of sign for the effective Newton's constant, has led to some discussion. In \cite{polem,polem1} it was noticed that for such a passage through the singularity some curvature invariants become infinite. In paper \cite{Bars1} a counter-argument was put forward. If one has enough conditions so as to match the nonsingular quantities before and after crossing the  singularities, then the singularities can be traversed. One can say that such a treatment of the problem of  singularities crossing is in some sense common to all the attempts to describe this process. A somewhat different approach to the problem of a cosmological singularity was developed in  papers \cite{Wetterich,Wetterich1}. There the author considers the so called variable gravity together  with the transitions between different frames and the reparametrization of the scalar field. In particular, he introduces a so-called freeze frame where the universe is very cold and slowly evolving. In the freeze picture the masses of elementary particles and the gravitational constant decrease with cosmic time, while  Newtonian attraction remains unchanged. The cosmological solution can be extrapolated to the infinite past in physical time - the universe has no beginning. In the equivalent, but singular,  Einstein frame cosmic history one finds the familiar Big Bang description.

The  conformal symmetry also was used in our approach, described in Sec. 5., where  we  implemented a particular choice of the coupling between the scalar field and the scalar curvature - the conformal coupling \cite{we-cross}. We  used a conformal coupling because in this case the relations between the parametrizations of the scalar field in the Jordan and in the Einstein frames have a simple explicit form.  Let us recapitulate the main idea of
  the paper \cite{we-cross}: when in the Einstein frame the universe arrives at the Big Bang - Big Crunch singularity, from the point of view of the evolution of its counterpart in the Jordan frame its geometry is regular, but the effective Planck mass has a zero value. The solution to the equations of motion in the Jordan frame is smooth at this point and on using the relations between the solutions of the cosmological equations in two frames one can describe the crossing of the cosmological singularity in a uniquely determined way. The contraction is replaced by the expansion (or vice versa)
and the universe enters into the antigravity regime. Analogously, when the geometry is singular in the Jordan frame it is regular in the Einstein frame, and using this regularity we can describe in a well determined way the crossing of the singularity in the Jordan frame. These scheme looks simpler than that, used in \cite{Bars,Bars1},
because only one scalar field is present. However, as we have shown in Sec. 6, such an approach is not suitable for the description of the singularity crossing in the anisotropic universes \cite{we-cross1}. Thus, we have suggested even more simple scheme, based on a particular reparametrization of the metric and of the scalar field for the description of the singularity crossing in the Bianchi-I universe \cite{we-cross1}. 

Concluding, we would like to say that while the idea of the crossing of the Big Bang - Big Crunch cosmological singularity looks rather exotic, it is not the most radical idea, discussed during the last decade. Indeed, we would like to mention the Conformal Cyclic Cosmology hypothesis by Roger Penrose \cite{Cycles}. According to this hypothesis the universe iterates through infinite cycles, with the future timelike infinity of each previous iteration being identified with the Big Bang singularity of the next. Thus, 
the past conformal boundary of one copy of a Friedmann-Lema\^{i}tre spacetime can be ``attached'' to the future conformal boundary of another, after an appropriate conformal rescaling. The elimination of the isotropic cosmological singularities by means of an infinite conformal rescaling was studied in detail in papers \cite{Tod,Tod1,Tod2}. Combining this method with the idea of the isotropization of an infinitely expanding universe one can arrive to an infinite sequence of aeons,
where the infinite expansion at the end of one aeon is transformed into the Big Bang at the beginning of another aeon.

\begin{acknowledgements}
This work was partially supported by the RFBR
Grant No. 17-02-01008.
\end{acknowledgements}

% BibTeX users please use one of
%\bibliographystyle{spbasic}      % basic style, author-year citations
%\bibliographystyle{spmpsci}      % mathematics and physical sciences
%\bibliographystyle{spphys}       % APS-like style for physics
%\bibliography{}   % name your BibTeX data base

% Non-BibTeX users please use

\end{document}